\magnification=1200
\def\tr{{\rm tr}}

\def\f#1#2{{\textstyle{#1\over #2}}}

\def\next{\hfil\break\noindent}
\def\R{{\bf R}}
\def\Quadrat#1#2{{\vcenter{\hrule height #2
  \hbox{\vrule width #2 height #1 \kern#1
    \vrule width #2}
  \hrule height #2}}}
\font\title=cmbx12
\hfuzz=2pt

{\title 
\centerline{Global dynamics of the Mixmaster model}}

\vskip 1cm

\noindent
{Alan D. Rendall}
\next
{Max-Planck-Institut f\"ur Gravitationsphysik}
\next
{Schlaatzweg 1}
\next
{14473 Potsdam}
\next
{Germany}

\vskip 1.5cm\noindent
{{\bf Abstract}} 

The asymptotic behaviour of vacuum Bianchi models of class A near the
initial singularity is studied, in an effort to confirm the standard
picture arising from heuristic and numerical approaches by mathematical
proofs. It is shown that for solutions of types other than VIII and IX
the singularity is velocity dominated and that the Kretschmann scalar
is unbounded there, except in the explicitly known cases where the 
spacetime can be smoothly extended through a Cauchy horizon. For types
VIII and IX it is shown that there are at most two possibilities for 
the evolution. When the first possibility is realized, and if the
spacetime is not one of the explicitly known solutions which can be 
smoothly extended through a Cauchy horizon, then there are infinitely
many oscillations near the singularity and the Kretschmann scalar is
unbounded there. The second possibility remains mysterious and it is
left open whether it ever occurs. It is also shown that any finite
sequence of distinct points generated by iterating the 
Belinskii-Khalatnikov-Lifschitz mapping can be realized approximately by a 
solution of the vacuum Einstein equations of Bianchi type IX.

\vskip 1cm\noindent
{\bf 1. Introduction}

Solutions of the vacuum Einstein equations with an $SU(2)$ isometry group
acting on spacelike hypersurfaces have been studied extensively over the
past twenty-five years or more. This class of spacetimes was christened the
\lq Mixmaster model\rq\ by Misner. A useful point of entry into the 
literature on the subject is the book of Hobill et. al. [11], where many of
the contributions are devoted to the Mixmaster model. Almost all the work
which has been done in this area is either heuristic in nature or based on
numerical calculations. There are very few rigorous results. Although the
numerical and heuristic approaches have led to differences of opinion, it
seems that by now a consensus has developed concerning various features of
the evolution of these spacetimes. This will be referred to in the following
as the \lq standard picture\rq. This paper is an investigation of what aspects
of this standard picture can be supported by rigorous theorems.

There are different motivations which explain the amount of effort which
has been put into understanding this very special class of spacetimes. One
of these is the desire to understand the nature of spacetime singularities
and in particular to find out whether the geometry near a singularity admits
a simple description or whether it is intrinsically complicated. The interest
of the Mixmaster model is that it is a case where although the setting is
relatively simple (the Einstein equations reduce to ordinary differential
equations) the behaviour of the solutions seems to be very complicated.
Another motivation for studying the Mixmaster model is provided by the
idea of Belinskii, Khalatnikov and Lifschitz [3] that it could provide
an approximate description for very general spacetime singularities. This
suggestion that very general spacetime singularities can be described
approximately by a relatively simple model, if true, is only useful insofar
as the model itself is understood. 

In this paper the ordinary differential equations which describe the 
Mixmaster model will be looked at from the point of view of the theory of
dynamical systems. This approach has been discussed extensively in the
book of Bogoyavlensky [5]. This book contains many interesting ideas,
but does not contain any general theorems on the Mixmaster model, the 
end result being an impressionistic informal description as to what the 
behaviour of generic solutions should be. Another way of writing the
equations for the Mixmaster model as a dynamical system has been presented
by Wainwright and Hsu [22]. Their equations cover a wider class of spacetimes
which are spatially homogeneous. In particular they include vacuum spacetimes
of Bianchi types of class A, i.e. Bianchi types I, II, VI${}_0$, VII${}_0$,
VIII and IX. This means considering spacetimes with three-dimensional isometry
groups more general than just $SU(2)$. The type IX spacetimes constitute the 
Mixmaster model. According to the standard picture general spacetimes of type 
VIII have  a complicated singularity similar to that of general spacetimes of 
type IX while the singularities in solutions of the other types are much 
simpler. It turns out to be very helpful to study all class A models in a 
unified way rather that trying to handle type IX in isolation. In [22] various
general properties of this system were determined, and in particular the 
nature of its critical points was investigated in detail. The work of 
Wainwright and Hsu is the starting point for the investigations of the 
present paper.

The results of the paper can be summarized as follows. Any solution
of the vacuum Einstein equations with a Bianchi symmetry of class A can be
classified, according to its behaviour near the initial singularity, into
one of three types. These will be referred to as \lq standard convergent\rq,
\lq standard oscillatory\rq\ and \lq anomalous\rq. (For the precise meaning
of this terminology, see the last paragraph of Section 4). Solutions of types 
I, II, VI${}_0$ and VII${}_0$ belong to the standard convergent type. This is 
also true of the NUT solutions of types VIII and IX, discussed further below.
All other solutions of types VIII and IX belong to either the standard
oscillatory or the anomalous type. If it could be shown that the anomalous
type never occurs then this would provide a strong confirmation of what was
referred to above as the standard picture. However, the results of this 
paper do not suffice to obtain that conclusion; in fact they do not even 
suffice to show that the standard oscillatory type ever occurs.

The detailed conclusions which can be made about the spacetimes of the 
different types just introduced concern on the one hand whether the 
singularity is a curvature singularity and on the other hand precise
statements about the convergent or oscillatory nature of the spacetimes
near the initial singularity. Those vacuum Bianchi models of class A which
admit an extension through a smooth Cauchy horizon have been determined
explicitly in [6]. Here it is shown that in any vacuum spacetime of Bianchi
class A which does not admit such an extension, and which is not anomalous,
the Kretschmann scalar 
$R_{\alpha\beta\gamma\delta}R^{\alpha\beta\gamma\delta}$
is unbounded in a neighbourhood of the initial singularity. This can be
seen as a result on the non-existence of \lq intermediate singularities\rq\ 
in this class of spacetimes. In general this result is weakened by the fact
that we have no control over how many anomalous spacetimes exist. On the
other hand, for types other than VIII and IX it provides optimal information.
As for the question of convergent or oscillatory behaviour, it is shown that
in the standard convergent case the singularity is velocity dominated. On
the other hand, in the standard oscillatory case it is shown that naturally
defined geometrical quantities undergo infinitely many oscillations as the
singularity is approached.

One aspect of the standard picture is that general solutions of the Mixmaster
model are supposed to be approximated in some sense near the singularity by
a mapping of the circle to itself, the BKL (Belinskii-Khalatnikov-Lifschitz) 
mapping. The results on oscillations mentioned up to now say nothing about 
this. However it can be proved to be true in a weak sense. The statement is 
that, given any finite sequence generated by the BKL map, there exists a 
solution of the Einstein vacuum equations of type IX (or indeed of type VIII) 
which reproduces this sequence with any desired degree of accuracy.

The paper is organized as follows. Section 2 presents some parts of the theory
of dynamical systems which will be necessary in the analysis. In Section 3 the
system of Wainwright and Hsu is recalled. Various results on the asymptotic
behaviour of a solution of the Wainwright-Hsu system as the singularity is
approached are proved in Section 4. In Section 5 these results are used to
obtain conclusions about curvature singularities and the convergent or
oscillatory behaviour of the corresponding spacetime. Moreover, it contains
the precise formulations of the results which have been presented informally 
in the preceding paragraphs.

\vskip .5cm\noindent
{\bf 2. Background on dynamical systems}

The purpose of this section is to collect together some facts on dynamical
systems which are useful in the analysis of the dynamics of Bianchi models.
For a more general introduction to relevant aspects of the theory the reader
is referred to [21]. Consider the system of ordinary 
differential equations:
$${dx\over dt}=f(x)\eqno(2.1)$$
where $f:\R^n\to \R^n$ is a $C^\infty$ mapping. (The choice of the 
differentiability class $C^\infty$ is not essential in what follows; it 
suffices for the applications in this paper.) The corresponding local flow 
is the mapping $F$ defined by the condition that $F(t,x)$ is the value at time
$t$ of the solution of (2.1) which takes on the value $x$ at time $t=0$,
provided the solution exists that long. A critical point of (2.1) is a
point $x$ with $f(x)=0$. An important element in the analysis of the global
properties of solutions of (2.1) is the study of the local behaviour of 
solutions near critical points. This can be described using the concept of
topological equivalence.

\vskip 10pt\noindent
{\bf Definition} If $x_0$ is a critical point of (2.1) and $y_0$ is a 
critical point of the system $dy/dt=g(y)$ with local flow $G(t,y)$, then the 
two systems are said to be topologically equivalent near the points $x_0$ and 
$y_0$ respectively if there exists a homeomorphism $\phi$ of an open 
neighbourhood $U$ of $x_0$ onto an open neighbourhood $V$ of $y_0$, with
$\phi(x_0)=y_0$ such that whenever $F(t,x)$ is defined for some $x\in U$, 
$G(t,\phi(x))$ is also defined, and is equal to $\phi(F(t,x))$.

\vskip 10pt\noindent
In other words, when expressed in appropriate local coordinates the solutions
of the two systems look identical. A critical point $x_0$ of the system (2.1) 
is called {\it hyperbolic} if the derivative $Df(x_0)$ of $f$ at that point 
(which defines the linearization of the system about that point) has no 
eigenvalues which are purely imaginary. A fundamental result on hyperbolic 
critical points is the Hartman-Grobman theorem ([9], p. 244):

\vskip 10pt\noindent
{\bf Theorem 2.1} If $x_0$ is a hyperbolic critical point of the system (2.1)
then the system is topologically equivalent near $x_0$ to the linearized
system $d\bar x/dt=Df(x_0)\bar x$ near the origin.

\vskip 10pt\noindent
Note that the homeomorphism whose existence is required in the
definition of topological equivalence cannot in general be chosen to
be a diffeomorphism.  However this fact will play no role in what
follows. Linear systems which have no purely imaginary eigenvalues can
be classified up to topological equivalence ([2], p. 48). Let $n_+$ and
$n_-$ be the number of eigenvalues with positive and negative real
parts respectively. Two linear systems of ODE without purely imaginary
eigenvalues are topologically equivalent near the origin if and only
if the corresponding values of $n_+$ and $n_-$ are equal. (In fact in
this case the equivalence is global, i.e. $U$ and $V$ can be taken to
be all of $\R^n$.) Thus up to topological equivalence, the only linear
systems on $\R^n$ with no purely imaginary eigenvalues are given by
the systems on $\R^{n_+}\times\R^{n_-}$ defined by $dy/dt=y$,
$dz/dt=-z$, with $y\in\R^{n_+}$ and $z\in\R^{n_-}$. A system of ODE of
this type is known as a standard saddle and there are only $n+1$
possibilities in $\R^n$. If $n_+$ or $n_-$ vanishes then there is not
a saddle in the usual sense, but rather a source or
sink. Nevertheless, we include this case in the definition of a
standard saddle. Combining this discussion with the Hartman-Grobman
theorem shows that in a neighbourhood of any hyperbolic critical point
the system (2.1) is topologically equivalent to a standard saddle.

What can be said in the case of a critical point $x_0$ which is not hyperbolic?
Let $E_+$, $E_0$ and $E_-$ be the spaces spanned by those generalized 
eigenvectors of $Df(x_0)$ whose real parts are positive, zero and negative
respectively. These are called the unstable, centre and stable subspaces
respectively. In general $\R^n$ is the direct sum of these three subspaces
and the hyperbolic case is that where the centre subspace reduces to zero.
Associated to these subspaces are locally invariant manifolds. A submanifold 
$M$ of an open neighbourhood $U$ of $x_0$ is called locally invariant if 
whenever $x\in M$ and $F(t,x)\in U$ then $F(t,x)\in M$.

\vskip 10pt\noindent
{\bf Definition} A (local) stable, centre or unstable manifold of the critical
point $x_0$ is a $C^1$ submanifold of an open neighbourhood $U$ of $x_0$ which
is locally invariant, which contains $x_0$, and whose tangent space at $x_0$
is the stable, centre or unstable subspace respectively.

\vskip 10pt\noindent
{}From the topological classification of hyperbolic fixed points discussed
above it is clear that a hyperbolic fixed point has stable and unstable
manifolds which are locally unique, i.e. the intersection of any such manifold
with a sufficiently small neighbourhood of the critical point is unique.
A generalization of this to non-hyperbolic critical points is given by the
centre manifold theorem [1]:

\vskip 10pt\noindent
{\bf Theorem 2.2} Let $x_0$ be a critical point of (2.1) and suppose that $f$
is $C^\infty$. Then there exist stable and unstable manifolds of class 
$C^\infty$ and a centre manifold of class $C^k$ for any finite $k$. The stable
and unstable manifolds are locally unique.

\vskip 10pt\noindent
The centre manifold need not be $C^\infty$ in general; the $C^k$ centre 
manifolds may shrink as $k$ increases. The centre manifold need also not be 
locally unique. (See [23], p. 210, for an example). Despite the non-uniqueness 
of the centre manifold, it can be used to formulate a generalization of the 
Hartman-Grobman theorem to non-hyperbolic critical points. This is the 
reduction theorem of Shoshitaishvili ([18], [19]). For a detailed proof
in English see [13].

\vskip 10pt\noindent
{\bf Theorem 2.3} Let $x_0$ be a critical point of (2.1). Then the system is 
topologically equivalent near $x_0$ to the Cartesian product of a standard 
saddle with the restriction of the flow to any centre manifold.
 
\vskip 10pt\noindent
Suppose for a moment that $f$ is such that all solutions of (2.1) can be 
extended so as to be defined for all real values of $t$. Then 
an $\alpha$-limit point of a solution $x(t)$ of the system (2.1)
is a point $x_*$ such that there exists a sequence $t_n$ with $t_n\to -\infty$
and $x(t_n)\to x_*$. The $\alpha$-limit set of the solution is the set of
all $\alpha$-limit points. The concepts of $\omega$-limit point and 
$\omega$-limit set are defined analogously by replacing $-\infty$ by $\infty$.
If not all solutions of (2.1) extend to global in time solutions it is possible
to rescale $f$ with a positive function so that all solutions of the rescaled
system do exist globally in time. This is discussed by Wainwright and Hsu[22],
section 3. The notions of $\alpha$- and $\omega$-limit
points can be applied to the rescaled system. The image of a solution of the 
original system in $\R^n$ is also the image of a solution of the rescaled
system. It is natural to define the $\alpha$- and $\omega$- limit points of
a solution $x(t)$ of the original system to be those of the corresponding
solution of the rescaled system. This definition can be rephrased in terms of
the original system. A solution $x(t)$ of (2.1) defined on some time interval
is said to be maximally extended if it cannot be extended to a solution on 
any strictly longer interval. If a maximally extended solution $x(t)$ is 
defined on the interval $(t_-,t_+)$, then the $\alpha$-limit and 
$\omega$-limit points of the solution are limits of sequences of the form 
$x(t_n)$ where $t_n\to t_-$ or $t_n\to t_+$. This is consistent with the 
previous definition in the case that the solution is defined for all real 
values of $t$.

Some standard properties of $\alpha$-limit sets will now be listed (cf. [23],
p. 46) Corresponding properties hold for $\omega$-limit sets. The 
$\alpha$-limit set of any solution is closed. It consists of a union of images
of solutions of the ODE. It is clear that it must also contain the 
$\alpha$-limit sets of these solutions. If the solution stays in a compact 
set for all $t<t_0$ then the $\alpha$-limit set is connected. Monotone 
functions are a useful tool for locating $\alpha$- and $\omega$-limit sets,
as shown by the following simple lemma.

\vskip 10pt\noindent
{\bf Lemma 2.1} Let $U$ be an open subset of $\R^n$ and let $F$ be a 
continuous function on $U$ such that $F(f(t))$ is strictly monotone for
any solution $f(t)$ of (2.1) as long as $f(t)$ is in in $U$. Then
no solution of (2.1) whose image is contained in $U$ has an $\alpha$- or 
$\omega$-limit point in $U$.

\noindent
{\bf Proof} Suppose that $p\in U$ is an $\alpha$-limit point of a solution
$f(t)$ whose image is contained in $U$. Then $F(f(t))$, being strictly
monotone, must have a limit, possibly infinite, as $t\to t_-$. On the other
hand, there is a sequence $t_n$ with $t_n\to t_-$ such that $F(f(t_n))$
coverges to $F(p)$. Hence $F(f(t))$ converges to $F(p)$. Thus $F$ is
constant on the $\alpha$-limit set of $f(t)$. There exists a solution 
$\bar f(t)$ which passes through $p$ and is entirely contained in the 
$\alpha$-limit set of $f(t)$. It follows from the above that $F$ is constant
along $\bar f(t)$, contradicting the property of strict monotonicity. The
argument for the $\omega$-limit set is similar.

\vskip .5cm\noindent
{\bf 3. The equations of Wainwright and Hsu}

There are many popular ways of writing the equations for Bianchi models.
The analysis which follows uses a form of the equations for models of
class A due to Wainwright and Hsu [22]. This is natural since it 
is in a sense an extension of their approach. The results of the present paper
concern vacuum spacetimes, whereas the Wainwright-Hsu system describes a
perfect fluid with linear equation of state $p=(\gamma-1)\rho$. However, 
vacuum models are described by the restriction to the submanifold defined
by the vacuum Hamiltonian constraint. The latter is independent of $\gamma$
and so the choice of $\gamma$ is immaterial for our purposes. However, the
explicit form of the equations off the constraint hypersurface is more or
less complicated according to the value of $\gamma$. In order to take 
advantage of this, $\gamma=2/3$ is chosen in the following. With this choice,
the equations are:
$$\eqalign{
N_1'&=(q-4\Sigma_+)N_1                     \cr
N_2'&=(q+2\Sigma_++2\sqrt{3}\Sigma_-)N_2    \cr
N_3'&=(q+2\Sigma_+-2\sqrt{3}\Sigma_-)N_3    \cr   
\Sigma'_+&=-(2-q)\Sigma_+-3S_+             \cr
\Sigma'_-&=-(2-q)\Sigma_--3S_-}\eqno(3.1)$$
where
$$\eqalign{
q&=2(\Sigma_+^2+\Sigma_-^2)                \cr
S_+&=\f12 [(N_2-N_3)^2-N_1(2N_1-N_2-N_3)]  \cr
S_-&=\f{\sqrt{3}}2 (N_3-N_2)(N_1-N_2-N_3)}\eqno(3.2)$$
and a prime denotes a derivative with respect to a certain time coordinate 
$\tau$. The (vacuum) Hamiltonian constraint is:
$$\Sigma_+^2+\Sigma_-^2+\f34 [N_1^2+N_2^2+N_3^2-2(N_1N_2+N_2N_3+N_3N_1)]=1
\eqno(3.3)$$
A solution of the vacuum Einstein equations with a Bianchi symmetry of
class A is described by a solution of (3.1) with initial data
satisfying (3.3). Then of course the whole solution lies in the
submanifold defined by (3.3). If $t$ is a Gaussian time coordinate
based on one of the orbits of the group action defining the symmetry,
and if $\tr k(t)$, denotes the mean curvature of these orbits, then
the time coordinate $\tau$ in (3.1) is defined by the relation
$\tau(t)=-\f13\int_{t_0}^t \tr k(t')dt'$. Since the determinant of the induced
metric of the orbits satisfies $d/dt(\log\det g)=-2\tr k$, $\tau^3$ is 
proportional to the volume form of the orbits. In other words $\tau$ represents
a length scale related to that volume form. For the precise definition of
the variables $(N_1,N_2,N_3,\Sigma_+,\Sigma_-)$ the reader is referred
to [22].  The variables $N_1$, $N_2$ and $N_3$ describe the
curvature of the spatial slices and, at the same, the Bianchi type. In
particular they allow all Bianchi types of class A to be included in a
single dynamical system. When a fluid is present, $\Sigma_+$ and
$\Sigma_-$ are related to the shear. In the vacuum case they represent
the trace-free part of the second fundamental form. All of these
variables are dimensionless, in the sense that if the spacetime metric
is multiplied by a constant they do not change. This means that they
are candidates for quantities which remain finite when a spacetime
singularity is approached. The system (2.3) has a three-fold symmetry,
which is not obvious when the equations are written in this form. It
is described in detail in [22]. In particular, it cyclically permutes
$N_1$, $N_2$ and $N_3$.

In discussing the global structure of the spacetimes considered in
this paper, we restrict to the maximal Cauchy development of data on a
spacelike orbit, which is in any case the only part of the spacetime
described directly by the variables in (3.1). It is well-known that a
vacuum Bianchi model of class A, and not of type IX, is geodesically
complete in one time direction which, reversing the time orientation
if necessary, we can assume to be the future. In the other time
direction there is a finite upper bound to the length of all causal
geodesics. In the case of Bianchi type IX, a bound of this kind holds
in both directions as a consequence of a result of Lin and
Wald[14,15]. In the latter case the variables in (3.1) are not
defined at the moment of maximum expansion. For the definition of
these variables involves dividing by the mean curvature $\tr k$, which
vanishes at that moment. Thus a maximally extended solution of (3.1)
of Bianchi type IX represents only half of a maximal Cauchy
development, where the model is either always expanding or always
contracting. Reversing the time orientation if necessary, it can be
assumed that it is expanding, so that the singularity lies in the
past.  With these conventions, if a maximal solution of (3.1) is
defined on the interval $(\tau_-,\tau_+)$ then the behaviour of the
corresponding spacetime near the singularity is encoded in the
behaviour of the solution as $\tau\to\tau_-$. For Bianchi types other than 
IX the behaviour as $\tau\to\tau_+$ encodes the behaviour of the spacetime in 
the time direction where it is complete. For Bianchi type IX it encodes the 
behaviour near the time of maximal expansion.

The Bianchi type to which a solution of (3.1) corresponds depends on
the values of $N_1$-$N_3$. If all three are zero the Bianchi type is
I. If precisely one is non-zero then it is II. If precisely two are
non-zero it is either VI${}_0$ (signs opposite) or VII${}_0$ (signs
equal). If all three are non-zero it is either IX (all signs equal) or
VIII (one sign different from the other two). The set of points
corresponding to any one of the Bianchi types is invariant under the
flow of the differential equation. A point of this set will be referred 
to as a point of the given Bianchi type. 

There are various monotone functions which can be defined. The function
$N_1N_2N_3$ is strictly monotone along any solution of type IX or VIII.
(Cf. the function $\Delta_1$ of [22].) For $(N_1N_2N_3)'=3q(N_1N_2N_3)$. This 
gives the result for $q\ne0$. If $q=0$ then the Hamiltonian constraint shows 
that either $\Sigma'_+$ or $\Sigma'_-$ is non-zero and this proves the desired
result. The functions 
$$Z_\epsilon=[\f43\Sigma_-^2+(N_2+\epsilon N_3)^2]/(-\epsilon N_2N_3)$$ 
are non-negative functions on the sets of points of type VI${}_0$ and VII${}_0$
with $N_1=0$ for $\epsilon=1$ and $\epsilon=-1$ respectively. They are 
strictly decreasing except when $\Sigma_-=0$. These statements are proved in 
[22]. The function $(1+\Sigma_+)^2$ is a non-increasing function along 
solutions on the union of points of types I, II, VI${}_0$ and VII${}_0$ with
$N_1=0$. For
$$d/d\tau [(1+\Sigma_+)^2]=-4(1+\Sigma_+)^2(1-\Sigma_+^2-\Sigma_-^2)
\eqno(3.4)$$
and the vacuum Hamiltonian constraint implies that $\Sigma_+^2+\Sigma_-^2\le 1$
for the given Bianchi types. This is analogous to a monotone function for
solutions of class B given by Hewitt and Wainwright [10]. 

In order to study the question of which of the spacetimes described by the
above equations have curvature singularities, it is useful to express
curvature invariants of the spacetimes in terms of the Wainwright-Hsu
variables.  Consider the example of the Kretschmann scalar 
$R_{\alpha\beta\gamma\delta}R^{\alpha\beta\gamma\delta}$. Define a 
dimensionless version of this quantity by
$$\kappa=R_{\alpha\beta\gamma\delta}R^{\alpha\beta\gamma\delta}/(\tr k)^4
\eqno(3.5)$$
Then $\kappa$ can be expressed as a polynomial in the variables
$N_1$, $N_2$, $N_3$, $\Sigma_+$ and $\Sigma_-$. This can conveniently be
done using the formulae for curvature given by Ellis and MacCallum [8].
Here only the expression for solutions of type I will be written out. In
that case:
$$\kappa=(-\f29+\f29\Sigma_++\f49\Sigma_+^2)^2+2(\f19\Sigma_+^2+\f13\Sigma_-^2
-\f19\Sigma_+-\f29)^2+\f2{27}\Sigma_-^2(2\Sigma_+-1)^2\eqno(3.6)$$
The fact that it is a positive definite expression is explained by the fact
that for a vacuum spacetime of Bianchi type I the magnetic part of the Weyl
tensor vanishes.

\vskip .5cm\noindent
{\bf 4. Possible limit sets}

It will now be shown among other things that the $\alpha$-limit set of
any solution of (3.1)-(3.3) of type I, II, VI${}_0$ or VII${}_0$ is a
single point of type I or type VII${}_0$. Consider first a solution of
type I. It corresponds to a critical point of (3.1) and so is its own
$\alpha$-limit set, as well as its own $\omega$-limit set. These
points form a circle, sometimes known as the Kasner ring. Next
consider a solution of type II. Using the threefold symmetry of the
equations, it may be assumed without loss of generality that it
satisfies $N_2=N_3=0$. Then an elementary calculation shows that
either $\Sigma_-=0$ or the ratio $(\Sigma_+-2)/\Sigma_-$ is time
independent. Taking account of the Hamiltonian constraint, this
reduces the motion to motion along a curve. The direction of motion
along this curve is controlled by the monotone function
$(1+\Sigma_+)^2$. It follows that the $\alpha$-limit set consists of
one point on the Kasner ring and the $\omega$-limit set of
another. There are three points on the Kasner ring which play a
special role. They are denoted by $T_1$, $T_2$ and $T_3$ in [22] and
have coordinates $(-1,0)$, $(1/2,\pm \sqrt{3}/2)$. They divide the
Kasner ring into three equal parts. The $\alpha$-limit set of a
solution of type II with $N_1\ne 0$ lies on the longer of the two open
arcs with endpoints $T_2$ and $T_3$, while the $\omega$-limit set lies
on the shorter of these arcs. The points $T_1$, $T_2$ and $T_3$ are
permuted cyclically by the threefold symmetry, which shows what
happens for type II solutions with $N_2\ne0$ or $N_3\ne0$. Associating
the $\alpha$-limit set of a type II solution with the $\omega$-limit
set of that solution defines a mapping from the Kasner ring with the
three exceptional points removed to the Kasner ring. This is a
realization of the famous BKL (Belinskii-Khalatnikov-Lifshitz) mapping.

Consider next solutions of type VI${}_0$ or VII${}_0$. Using the threefold
symmetry, it can be assumed without loss of generality that $N_1=0$. Equation
(3.4) and Lemma 2.1 imply that for any $\alpha$-limit point 
$\Sigma_+^2+\Sigma_-^2=1$ or $\Sigma_+=-1$. For these solutions the 
Hamiltonian constraint reduces to:
$$\Sigma_+^2+\Sigma_-^2+\f34 (N_2-N_3)^2=1\eqno(4.1)$$
and so if $\Sigma_+=-1$ then $\Sigma_+^2+\Sigma_-^2=1$ anyway. Moreover
$N_2=N_3$. This means in particular that there
are no $\alpha$-limit points of type VI${}_0$. The $\alpha$-limit set 
must contain a solution of the equations, which satisfies these conditions
everywhere. The second and third equations of (3.1) then give either 
$N_2=N_3=0$ or $\Sigma_-=0$, whence $\Sigma_+=\pm 1$. Where $\Sigma_+=-1$, 
$\Sigma_++1=0$ and since this function is non-negative and non-increasing 
along solutions, it follows that any solution which has a point of this type 
as an $\alpha$-limit point must be a time independent solution. Apart from 
these time-independent solutions, the remaining possibilities are that either 
the $\alpha$-limit set contains a point with $\Sigma_+=1$, $\Sigma_-=0$ and 
$N_2=N_3$, or that the entire $\alpha$-limit set is contained in the Kasner 
ring. The points in the first of these two cases constitute the image of
a whole solution and so if one of them belongs to the $\alpha$-limit set of
a solution, they all do. In particular, this would mean that the solution
would not remain in a compact set. It will now be seen that this is impossible.
In the Bianchi VI${}_0$ case this is simple, since the Hamiltonian constraint 
shows that the set of Bianchi VI${}_0$ points is compact. The Bianchi 
VII${}_0$ case requires a little more work. Consider the behaviour of the
non-increasing function $Z_{-1}$ along any solution of type VII${}_0$.
Either it is identically zero or or it is bounded below by a positive constant
for all $\tau\le\tau_0$ and some fixed $\tau_0$. In the first case 
$N_2=N_3$, $\Sigma_-=0$ and $\Sigma_+=\pm1$. Then $N_2'=2(1+\Sigma_+)N_2$ and
the solution remains in a compact set as $\tau\to -\tau_-$. In the second case,
for any fixed $\tau_0$,
$$N_2N_3\le C[\f43\Sigma_-^2+(N_2-N_3)^2]\eqno(4.2)$$
holds for all $\tau\le\tau_0$. Combining this with the Hamiltonian constraint 
(4.1) proves that the solution remains in a compact set as $\tau\to -\tau_-$.
The fact that solutions of these types remain in a compact set for all 
sufficiently negative times also proves that their $\alpha$-limit sets are
non-empty. 

The points $T_1$, $T_2$ and $T_3$ of the Kasner ring correspond to spacetimes
obtained by making identifications in a subset of Minkowski space. In 
particular they are flat. In fact they are the only flat Kasner solutions,
since from (3.6) the Kretschmann scalar is non-zero at all other points of
the Kasner ring. The points of type of type VII${}_0$ which satisfy 
$\Sigma_+=-1$ and $\Sigma_-=0$ represent the same spacetime as the point
$T_1$. The reason for this is that the Wainwright-Hsu variables depend not
only on the spacetime but on a choice of frame. The flat Kasner solutions
admit a frame of type I and a one-parameter family of frames of type VII${}_0$,
which can all be used. Thus one spacetime is represented by different points
in $\R^5$. These type VII${}_0$ solutions and those obtained from them by
applying the threefold symmetry will be referred to as the flat type VII${}_0$
solutions. This terminology will be justified later, when it is shown that 
they are the only flat solutions of type VII${}_0$.

It has now been shown that, except for the flat solutions of type VII${}_0$,
the $\alpha$-limit set of any solution of type I, II, VI${}_0$ or VII${}_0$ 
is contained in the Kasner ring. It will now be shown that it consists of
a single point of the Kasner ring. To see this, consider the behaviour of
the function $(1+\Sigma_+)^2$ along the given solution. It cannot take two
different values on the $\alpha$-limit set, since then it could not be 
monotonic along the solution. Thus the $\alpha$-limit set of any 
given solution consists of points on the Kasner ring where $\Sigma_+$ takes on
the same value. But for any given value of $\Sigma_+$ there is only one
point, in which case the desired result follows, or there are two. In the
latter case it follows from the connectedness of the $\alpha$-limit set
that only one of the two can be contained in it. The following theorem has now
been proved: 

\vskip 10pt\noindent
{\bf Theorem 4.1} The $\alpha$-limit set of a solution of (3.1)-(3.3) of type
I, II, VI${}_0$ or VII${}_0$ is a single point of type I or a flat
point of type VII${}_0$, the latter only being possible if the solution is
time independent.

\vskip 10pt\noindent
Note that all the arguments used to prove this were elementary, and that the
reduction theorem (Theorem 2.3) was not used. In studying the $\alpha$-limit
sets of solutions of types VIII and IX the heavier machinery is required.
Note first that, due to the strictly monotone function for these Bianchi
types presented in Section 3, the $\alpha$-limit set of any solution of type
VIII or IX is contained in the set $N_1N_2N_3=0$, which consists of
points of the other, simpler, Bianchi types. It follows that the $\alpha$-limit
set, if non-empty, contains the image of a solution of one of these types. It
then also contains the $\alpha$-limit set of that solution. Thus it follows
from Theorem 4.1 that:

\vskip 10pt\noindent
{\bf Theorem 4.2} The $\alpha$-limit set of a solution of (3.1)-(3.3) of type
VIII or IX is either empty or contains a point of type I or 
contains a flat point of type VII${}_0$.

\vskip 10pt\noindent
In the rest of this section it will be shown that, except for a small set of
well-understood solutions, if the $\alpha$-limit set of a solution of type
VIII or IX is non-empty, then it either contains a non-empty open subset of
flat points of type VII${}_0$ or a non-flat point of the Kasner ring and that 
in the latter case it contains more than one non-flat point of the 
Kasner ring.

\vskip 10pt\noindent
{\bf Theorem 4.3} The $\alpha$-limit set of a solution of (3.1)-(3.3) of type
VIII or IX has at least one of the following properties:
\next
(i) it is empty
\next
(ii) it contains at least two distinct points of type I, at least one of which
is non-flat
\next
(iii) it consists of flat points of type I or VII${}_0$ and contains a 
non-empty  open subset of the set of points of type VII${}_0$.
\next
(iv) it is a flat point of type I or VII${}_0$ and the solution is contained
in the unstable manifold of that point 

\noindent
{\bf Proof} The essential point is to apply the reduction theorem (Theorem 2.3)
to the critical points of the system. In this context it is important to note
that the Kasner ring is a centre manifold for each of the non-flat points 
contained in it, while the manifolds $N_1=0$, $N_2=0$ and $N_3=0$ are centre
manifolds for the flat solutions of type I and VII${}_0$ which they contain.
This follows from the computation of the eigenvalues and eigenspaces of the
critical points carried out by Wainwright and Hsu [22] and the obvious fact
that the manifolds in question are invariant manifolds.

Consider now a solution of type VIII or IX with non-empty $\alpha$-limit set.
First it will be shown that if its $\alpha$-limit set contains one non-flat
point $p$ of type I, it contains at least two such points. The Kasner ring,
which is a centre manifold for $p$, consists of critical points. Hence, by the
reduction theorem, a neighbourhood of $p$ is foliated by invariant manifolds
of codimension one, each of which meets the Kasner ring in a single point. In
each of these submanifolds the flow is topologically equivalent to a standard
saddle. It follows that if the solution converges to $p$ as $\tau\to\tau_-$,
it must lie on the unstable manifold of $p$ and that otherwise the 
$\alpha$-limit set must contain the stable manifold. Suppose without loss
of generality that $p$ belongs to the shorter of the two arcs of the Kasner
ring joining $T_2$ to $T_3$. Then the analysis of the linearization in [22]
shows that the subspaces $N_2=N_3=0$ and $N_1=0$ are the stable and unstable 
manifolds respectively. Since the unstable manifold contains no points of
types VIII or IX, it follows that the $\alpha$-limit set of the solution
under consideration contains the stable manifold, which is the image of a
solution of type II. Since the $\alpha$-limit set is closed, it follows that
it must also contain the image of $p$ under the BKL map. Thus it contains at
least two distinct points.

Next the case will be considered where no $\alpha$-limit point of the
solution is a non-flat point of type I. Because of the monotone function
$N_1N_2N_3$, all $\alpha$-limit points must satisfy $N_1=0$, $N_2=0$ or
$N_3=0$. A $\alpha$-limit point of type II or type VI${}_0$ or a non-flat 
point of type I or VII${}_0$ is not possible, as follows from
Theorem 4.1. Hence the entire $\alpha$-limit set consists of flat points
of type I or VII${}_0$. Note also that no solution on the centre manifold
can approach one of these points as $\tau\to\tau_-$, unless it is the
solution which stays at that point all the time. Hence, by the reduction
theorem,  the solution under
consideration can only approach a point of that type if it lies on the 
stable manifold of that point. Using the calculations of [22] once more,
the stable manifold can be identified. After applying the threefold symmetry
if necessary, it is given by the equations $N_2=N_3$, $\Sigma_-=0$. The Bianchi
type IX solutions satifying these conditions are the Taub-NUT solutions. 
Similarly, solutions of type VIII (or indeed of type II) satisfying these
conditions are the NUT solutions of those types described by Siklos[20]. It
remains to consider the case of a solution whose $\alpha$-limit points are
flat points of type I or VII${}_0$, but which does not converge to such a
point. By the reduction theorem, a solution of this type which has $p$ as
an $\alpha$-limit point must repeatedly leave the neighbourhood of $p$
whose existence is asserted by the theorem. In particular, it crosses a
sphere of any sufficiently small radius about $p$ infinitely many times.
Thus some $\alpha$-limit point lies on any such sphere. It follows that
the $\alpha$-limit set contains an open subset of the set of flat points of 
type VII${}_0$. 

\vskip 10pt\noindent
The notions of \lq standard convergent\rq, \lq standard oscillatory\rq\ and
\lq anomalous\rq\ mentioned in the introduction will now be defined in terms
of the four cases of the theorem. A solution of type VIII or IX is standard
convergent if it satisfies property (iv) of Theorem 4.3. A solution of any 
other Bianchi type of class A is always standard convergent. (This terminology
is justified by Theorem 4.1.) A solution is standard oscillatory if it 
satisfies property (ii) of Theorem 4.3. A solution which is neither standard 
convergent nor standard oscillatory is called anomalous.

\vskip .5cm\noindent
{\bf 5. Conclusions}

The results of the previous section on the $\alpha$-limit sets of solutions
of (3.1)-(3.3) will now be interpreted in terms of properties of the
corresponding spacetimes. The first issue to be considered is that of
curvature singularities. Here the theorems of the last section give 
complete information for all types other than VIII and IX. For these two
types the information obtained is less satisfactory.

\vskip 10pt\noindent
{\bf Theorem 5.1} Let $(M,g)$ be a vacuum spacetime with a Bianchi symmetry
of class A. Suppose that the time orientation has been chosen such that 
the maximal Cauchy development of data on a homogeneous hypersurface is
past incomplete, so that there is a past singularity. Then at least one of
the following holds:

\noindent
(i) the Kretschmann scalar is unbounded in a neighbourhood of the initial
singularity
\next
(ii) the maximal Cauchy development can be extended through a smooth Cauchy
horizon, and then the spacetime is flat or a NUT spacetime
\next
(iii) the corresponding solution has property (i) or property (iii) of
Theorem 4.3

\noindent
{\bf Proof} Assume that property (iii) of the conclusions of this theorem does
not hold. Then if the solution of (3.1)-(3.3) corresponding to the given 
spacetime is of type VIII or IX, it has property (ii) or (iv) of the
conclusions of Theorem 4.3. The solutions which have property (iv) of Theorem 
4.3 have already been identified; they are the NUT spacetimes. It is known 
that they admit an extension through a smooth Cauchy horizon (cf. [6]). Thus 
it remains to prove the theorem in the case that a solution of type VIII or IX
has property (ii) of Theorem 4.3 and in the case that the Bianchi type is 
neither VIII nor IX. {}From Theorems 4.1 and 4.3 it follows that either the
spacetime corresponds to a flat point of type I or VII${}_0$
or that there is a non-flat point of type I in the 
$\alpha$-limit set. In the first of these cases, it is known that the 
spacetime can be extended through a smooth Cauchy horizon [6]. In the 
second case the values of the scalar $\kappa$ introduced in Section 2
have a cluster point $\kappa_0\ne 0$ as $\tau\to\tau_-$, where $\kappa_0$
is the value of $\kappa$ at a non-flat $\alpha$-limit point of the solution
on the Kasner ring. Since the Kretschmann scalar is equal to
$\kappa(\tr k)^4$, it follows that $\limsup R^{\alpha\beta\gamma\delta}  
R_{\alpha\beta\gamma\delta}=\infty$. This completes the proof.

\vskip 10pt\noindent
According the standard picture of the Mixmaster solutions, case (iii) of 
Theorem 5.1 should not occur. If that were the case, a clean result would be 
obtained on the non-existence of \lq intermediate singularities\rq. This 
would mean that every solution possessed either a smooth Cauchy horizon or an 
unbounded curvature invariant near the singularity. Note that this conclusion 
has been proved for solutions of type other than VIII and IX, as can be seen 
from an examination of the above proof. In fact, slightly more has been proved,
since it has not only been shown that the Kretschmann scalar is unbounded near
the singularity, but also that it tends to infinity there at the same rate as
rate $(\tr k)^4$. Computing curvature invariants other than the Kretschmann 
scalar would not lead to an improved result for types VIII and IX using the 
above techniques, since all curvature invariants vanish at the flat points.  

The other important point of interpretation concerns the question of convergent
or oscillatory behaviour near the singularity, and velocity dominance. The
notion of a velocity dominated singularity was introduced by Eardley, Liang  
and Sachs [7] and a related notion was used by Isenberg and Moncrief [12] 
(see also [16], [17] for other applications). The general idea is that near 
the singularity the spacetime should be approximated in some appropriate sense
by a solution of Bianchi type I. In the present context it might be tempting to
say that a solution of (3.1)-(3.3) was velocity dominated if its $\alpha$-limit
set consisted of a point on the Kasner ring. However, this is too simple, since
the flat points of type VII${}_0$ represent the same spacetimes as the flat 
points of Bianchi type I. Thus the following definition will be used:

\vskip 10pt\noindent
{\bf Definition} A singularity in a spacetime with a given 
3+1-decomposition is called {\it weakly velocity dominated} if the generalized
Kasner exponents $p_i(t,x)$ converge to a limit for each fixed $x$ as the 
singularity is approached and if their limits satisfy $\Sigma_i p_i^2=1$.

\vskip 10pt\noindent
This definition requires some further explanation. Suppose that the 
hypersurfaces of constant $t$ have nowhere vanishing mean curvature near the
singularity. (If this is not the case, the definition is deemed to be 
violated) Let $\lambda_i$ be the eigenvalues of the second fundamental form.
Then the mean curvature is $\tr k=\Sigma_i \lambda_i$ and the generalized
Kasner exponents are defined to be $p_i=\lambda_i/(\tr k)$. In a Kasner
solution they are constants and satisfy the equation required in the above
definition. For the Bianchi spacetimes considered here, the definition will
be applied with a 3+1-decomposition defined by Gaussian coordinates based
on a homogeneous hypersurface. Then the generalized Kasner exponents are
linear combinations of $\Sigma_-$ and $\Sigma_+$ and the equation of the
definition becomes $\Sigma_+^2+\Sigma_-^2=1$. It is easy to think up 
alternative definitions. For instance, the condition $\Sigma_i p_i^2=1$
could be replaced by the condition $\rho_i/(\tr k)^2\to 0$, where $\rho_i$
are the eigenvalues of the spatial Ricci tensor. The resulting definition 
is a priori stronger than that given above but in all the cases where
the above definition is shown to be satisfied in the following the stronger
definition is also satsisfied. 

Returning to the question of oscillatory behaviour, a function $F$ on an
interval $(\tau_-,\tau_+)$ will be said to have infinitely many oscillations
as $\tau\to\tau_-$ if there exist two numbers $a$, $b$ with $a<b$ and a
sequence $\tau_n$ with $\tau_n\to\tau_-$ as $n\to\infty$ such that for
any positive integer $k$, $F(\tau_{2k-1})\le a$ and $F(\tau_{2k})\ge b$.

\vskip 10pt\noindent
{\bf Theorem 5.2} Let $(M,g)$ be a vacuum spacetime with a Bianchi symmetry
of class A. Suppose that the time orientation has been chosen such that 
the maximal Cauchy development of data on a homogeneous hypersurface is
past incomplete, so that there is a past singularity. Then either:

\noindent
(i) the singularity is weakly velocity dominated
\next
(ii) the spacetime is of type VIII or IX and $\Sigma_+$ or $\Sigma_-$ has
infinitely many oscillations as the singularity is approached.
\next
(iii) the spacetime is of type VIII or IX and the corresponding solution of 
(3.1)-(3.3) fails to be contained in a compact set as $\tau\to\tau_-$.  

\vskip 10pt\noindent
{\bf Proof} If the spacetime is not of type VIII or IX then the result
follows immediately from Theorem 4.1. Consider a spacetime of type VIII or IX
such the corresponding solution of (3.1)-(3.3) is contained in a compact set 
as $\tau\to\tau_-$. Then it must have one of the properties (ii), (iii) or 
(iv) of Theorem 4.3. If it has property (iv) of that theorem the singularity 
is weakly velocity dominated. If it has property (ii) it repeatedly comes 
close to two different points of the Kasner ring as $\tau\to\tau_-$ and so 
$\Sigma_+$ or $\Sigma_-$ has infinitely many oscillations as the singularity 
is approached. If it does not belong to any of these cases of Theorem 4.2 
then its $\alpha$-limit set must be a compact subset of the set of flat points
of type I and VII${}_0$. In that case $\Sigma_+\to -1$ and $\Sigma_-\to 0$
and the solution is weakly velocity dominated.

\vskip 10pt\noindent
This result implies in particular that solutions which are neither of type 
VIII or IX are (weakly) velocity dominated. The standard picture indicates 
that the only solutions of types VIII and IX which are velocity dominated are 
the NUT solutions and that case (iii) of Theorem 5.1 would be superfluous. 
However, it has not been proved that this is the case. 

\vskip 10pt\noindent
Although the BKL mapping plays an implicit role in these results, they do
not provide any direct confirmation of the idea, which is part of the 
standard picture, that the BKL mapping represents an approximation to solutions
of the Mixmaster model in some sense. The next result is a statement which 
goes in that direction. Recall that the BKL mapping is a mapping from the
Kasner ring with the three exceptional points $T_1$, $T_2$ and $T_3$ removed
to the Kasner ring. Starting with a point of the Kasner ring and applying
the BKL mapping repeatedly produces a sequence of points. If one of these
points is one of $T_1$, $T_2$ or $T_3$ then the map can no longer be applied
and only a finite sequence can be defined. If, on the other hand, this never
happens, an infinite sequence is obtained. This may or may not be periodic.
By a finite non-repeating sequence of BKL iterates we mean a finite initial 
piece of the iteration, which never hits the same point twice, whether or not
the sequence as a whole is finite or infinite for that starting point. In the
case that the whole sequence is periodic, we need to cut it off after some
number of iterations smaller than the period. To measure the quality of the 
approximation the Euclidean metric in the space $\R^5$ where the 
Wainwright-Hsu system is defined will be used. The distance between points 
$x$ and $y$ in this metric will be denoted by $d(x,y)$.

\vskip 10pt\noindent
{\bf Theorem 5.3} Let $\{x_1,\ldots,x_n\}$ be a finite sequence of BKL iterates
and $\epsilon>0$. Then there exists a $\delta>0$ and a solution of the vacuum 
Einstein equations of Bianchi type IX which, when written in the 
Wainwright-Hsu variables as $f(\tau)$, has the following properties:

\noindent
1. $\delta<\epsilon$ and the balls of radius $\delta$ about each of the 
points $x_i$ have disjoint closures.
\next
2. There exists a finite sequence of times $\tau_1>\ldots >\tau_n$
such that $f(\tau_i)$ is contained in the open ball of radius $\delta$ about 
$x_i$ while for $\tau_i\le\tau\le\tau_{i+1}$ the solution does not come
closer than $\delta$ to any $x_j$ other than $x_i$ or $x_{i+1}$. 

\noindent
{\bf Proof} As explained in Section 3 each iteration of a point via the BKL
map can be represented by a solution of Bianchi type II. For $i=1,2,\ldots,
n-1$, let $s_i$ be the image of the Bianchi II solution which produces 
$x_{i+1}$ from $x_i$. Choose some $\eta>0$ smaller than $\epsilon$ such
that the closures of the balls $B_i$ of radius $\eta$ about the points $x_i$
are disjoint and such that on each of these balls the flow has the local
product structure of the reduction theorem. Now choose $\delta_0<\eta$ 
such that any point which is a distance less than $\delta_0$ from each of two 
distinct $s_i$ must be inside one of the balls $B_i$. That such a $\delta_0$ 
exists follows from the fact that the intersections of each $s_i$ with the 
complement of the union of the $B_j$ are disjoint compact sets. Let $S$ be 
the set of all points which are a distance less than $\delta_0$ from some 
$s_i$. It is an open set which contains all $x_i$. For $1\le i\le n-1$ let 
$y_i$ be a point of $s_i$ which is a distance less than $\delta_0$ from $x_i$.
For $2\le i\le n$ let $z_i$ be a point of $s_{i-1}$ which a distance less 
than $\delta_0$ from $x_i$. Let $\delta_1$ be such that the ball of radius 
$\delta_1$ about $y_1$ is contained in the ball of radius $\delta_0$ about 
$x_1$. There exists a time $\Delta\tau_1$ such that the solution with 
$f(\tau)=z_2$ satisfies $f(\tau+\Delta\tau_1)=y_1$. Choose $\epsilon_2>0$ 
such that if $d(x,z_2)<\epsilon_2$ then $d(F(\Delta\tau_1,x),y_1)<\delta_1$.
Here $F$ denotes the local flow of the differential equation. 
By reducing the size of $\epsilon_2$ if necessary, it can be ensured that if
$d(x,z_2)<\epsilon_2$, then $F(\tau,x)\in S\backslash B'_i$ for all 
$\tau\le\Delta\tau_1$, where $B_i'$ is the complement of $\bar B_i\cup\bar 
B_{i+1}$. Now choose $\delta_2$ such any $x$ with the correct signs of $N_1$, 
$N_2$ and $N_3$ which lies in the open ball of radius $\delta_2$ about $y_2$ 
passes through the open ball of radius $\epsilon_2$ about $z_2$ and stays in 
$S$ as long as it stays in the ball of radius $\delta_0$ about $x_2$. The 
existence of a number $\delta_2$ with this property follows from the local 
saddle point structure of the flow near the point $x_2$. In a similar way we 
can recursively define $\Delta\tau_i$ for $i=2,\ldots,n-1$, $\epsilon_i$
for $i=3,\ldots,n$ and $\delta_i$ for $i=3,\ldots,n-1$. Now let 
$\delta=\delta_{n-1}$ and let $x_*$ be any point of type IX such that 
$d(x_*,z_n)<\delta$. We claim that the solution with initial data $x_*$ for
$\tau=0$ has the desired  properties. Note first that by construction it 
enters each of the balls $B_i$ and is contained in $S$. Let $\tau_n=0$.
Let $\sigma_{n-1}=\tau_n+\Delta\tau_{n-1}$. Let $\tau_{n-1}>\sigma_{n-1}$ be a 
time such that $d(f(\tau_{n-1}),z_{n-1})<\epsilon_{n-1}$ and $f(\tau)$ is
in $B_{i-1}$ for all $\tau$ in the interval $[\sigma_{i-1},\tau_{i-1}]$.
Continuing in this way, the other $\tau_i$ can be defined recursively.  

\vskip 10pt\noindent
The intuitive meaning of this theorem is that the solution visits small
neighbourhoods of the points generated by th BKL iteration and, moreover,
does so in the order determined by the iteration.
The analogous theorem holds with type VIII replaced by type IX, the proof 
being almost identical. This result shows that arbitrarily long but finite
segments of a BKL iteration can be realized by a solution of the Einstein
equations. However, it says nothing about the full iteration, if it happens
to be infinite. This is natural, since for a given measure of error $\epsilon$
and a given starting point specified with precision $\epsilon$ it cannot be
expected that the qualitative behaviour of the sequence of iterates is 
determined.

What do Theorems 5.1-5.3 tell us about Mixmaster dynamics? They tell us that
a finite sequence of BKL iterates corresponds to a solution of the exact
equations. They do not tell us the ultimate fate of the solution at early times
because the possibility is left open that eventually all solutions show the
behaviour described above as anomalous. Numerical results make this 
possibility implausible. Recent calculations[4] have been able to observe 150
BKL iterates in numerical solutions of the Mixmaster model. Moreover, there is
no indication that things change at that point and anomalous behaviour seems
never to have been observed numerically. Thus it seems reasonable to suppose
that at least a large open set of initial data for the Mixmaster model gives
rise to standard oscillatory behaviour. On the other hand, it it difficult to
see how the mathematical techniques used in this paper could lead to a proof
of this fact. The key question is, what other techniques might do so.

\vskip .5cm\noindent
{\bf Acknowledgements} I am grateful to J. Isenberg and J. Wainwright for
helpful discussions and correspondence.

\vskip .5cm\noindent
{\bf References}

\vskip 10pt\noindent
[1] Abraham R and Robbin J 1967 {\it Transversal mappings and flows} (New York:
Benjamin)
\next
[2] Arnold V I and Ilyashenko Yu S 1988 {\it Dynamical Systems I} ed D V Anosov
and V I Arnold (Berlin: Springer)
\next
[3] Belinskii V A, Khalatnikov I M and Lifshitz E M 1982 A general solution of
the Einstein equations with a time singularity. {\it Adv. Phys.}
{\bf 31} 639-667
\next
[4] Berger B K, Garfinkle D and Strasser E 1996 New algorithm for Mixmaster 
dynamics. Preprint gr-qc/9609072
\next
[5] Bogoyavlensky O I 1985 {\it Qualitative theory of dynamical systems in 
astrophysics and gas dynamics} (Berlin: Springer)
\next
[6] Chru\'sciel P T and Rendall A D 1995 Strong cosmic censorship in vacuum 
spacetimes with compact locally homogeneous hypersurfaces. {\it Ann. Phys.} 
{\bf 242} 349-385
\next
[7] Eardley D, Liang E. and Sachs R 1972 Velocity-dominated singularities in 
irrotational dust cosmologies {\it J. Math. Phys.} {\bf 13} 99-106 
\next
[8] Ellis G and MacCallum M 1969 A class of homogeneous cosmological models
{\it Commun. Math. Phys.} {\bf 12} 108-141
\next
[9] Hartman P 1982 {\it Ordinary Differential Equations} (Boston: Birkh\"auser)
\next
[10] Hewitt C and Wainwright J 1993 A dynamical systems approach to Bianchi 
cosmologies: orthogonal models of class B {\it Class. Quantum Grav.} {\bf 10} 
99-124
\next
[11] Hobill D, Burd A and Coley A 1994 {\it Deterministic Chaos in General 
Relativity}   (New York: Plenum)
\next
[12] Isenberg J and Moncrief V 1990 Asymptotic behaviour of the 
gravitational field and the nature of singularities in Gowdy 
spacetimes. {\it Ann. Phys.} {\bf 199} 84-122
\next
[13] Kirchgraber U and Palmer K J 1990 {\it Geometry in the neighborhood of 
invariant manifolds of maps and flows and linearization} (Harlow: Longman)
\next
[14] Lin X-F and Wald R 1989 Proof of the closed universe recollapse 
conjecture for diagonal Bianchi type IX cosmologies {\it Phys. Rev. D} 
{\bf 40} 3280-3286
\next
[15] Lin X-F and Wald R 1990 Proof of the closed universe recollapse 
conjecture for general Bianchi type IX cosmologies {\it Phys. Rev. D} 
{\bf 41} 2444-2448
\next
[16] Rein G 1996 Cosmological solutions of the Vlasov-Einstein system with 
spherical, plane and hyperbolic symmetry {\it Math. Proc. Camb. Phil. Soc.} 
{\bf 119} 739-762
\next
[17] Rendall A D 1995 On the nature of singularities in plane symmetric 
scalar field cosmologies {\it Gen. Rel. Grav.} {\bf 27} 213-221
\next
[18] Shoshitaishvili A N 1972 Bifurcations of topological type at singular 
points of paramet\-rized vector fields {\it Funct. Analysis Applications} 
{\bf 6} 169-170 
\next
[19] Shoshitaishvili A N 1975 Bifurcations of topological type at singular 
points of paramet\-rized vector fields {\it Tr. Semin. I. G. Petrovskii} 
{\bf 1} 279-309 (Russian)
\next
[20] Siklos S T C 1976 Two completely singularity-free NUT spacetimes. {\it 
Phys. Lett. A} {\bf 59} 173-174
\next
[21] Wainwright J 1994 {\it Deterministic Chaos in General Relativity} ed 
D Hobill, A Burd and A Coley  (New York: Plenum)
\next
[22] Wainwright J and Hsu L 1989 A dynamical systems approach to Bianchi 
cosmologies: orthogonal models of class A. {\it Class. Quantum Grav.} {\bf 6} 
1409-1431
\next
[23] Wiggins S 1990 {\it Introduction to applied nonlinear dynamical systems 
and chaos.} (Berlin: Springer)

\end